\documentclass[%
 reprint,
 superscriptaddress,
 amsmath,amssymb,
 aps,
 nature,
]{revtex4-1}

\usepackage{graphicx}% Include figure files
\usepackage{dcolumn}% Align table columns on decimal point
\usepackage{bm}% bold math
\usepackage{xcolor}
\usepackage{algorithm}
\usepackage{algpseudocode}
\usepackage{url} 
\usepackage{makecell}
\usepackage{multirow}
\usepackage{booktabs}
\usepackage{graphicx}
\usepackage{subfig}
\usepackage{comment}
\usepackage{braket}
\usepackage[amssymb]{SIunits}
\usepackage{booktabs,tabularx}

\DeclareMathOperator\erf{erf}
\DeclareMathOperator\erfc{erfc}

\def\br{{\bm r}}

\def\phib{\phi}
\def\vi{v}

\def\br{{\bm r}}
\def\bR{{\bm R}}

\newcommand {\pder}[2]{\frac{\partial  #1}{\partial #2} }
\newcommand {\vder}[2]{\frac{\delta  #1}{\delta #2} }

\begin{document}

%\title{Charge-dependent Atomic Cluster Expansion}
\title{Charge-constrained Atomic Cluster Expansion}

\author{Matteo Rinaldi}
\email{matteo.rinaldi@rub.de}
\affiliation{Interdisciplinary Centre for Advanced Materials Simulation, Ruhr-Universit{\"a}t Bochum, 44801 Bochum, Germany}
\author{Anton Bochkarev}
\email{anton.bochkarev@rub.de}
\affiliation{Interdisciplinary Centre for Advanced Materials Simulation, Ruhr-Universit{\"a}t Bochum, 44801 Bochum, Germany}
\author{Yury Lysogorskiy}
\email{yury.lysogorskiy@rub.de}
\affiliation{Interdisciplinary Centre for Advanced Materials Simulation, Ruhr-Universit{\"a}t Bochum, 44801 Bochum, Germany}
\author{Ralf Drautz}
\email{ralf.drautz@rub.de}
\affiliation{Interdisciplinary Centre for Advanced Materials Simulation, Ruhr-Universit{\"a}t Bochum, 44801 Bochum, Germany}

\date{\today}

\begin{abstract}
The atomic cluster expansion (ACE) efficiently parameterizes complex energy surfaces of pure elements and alloys. Due to the local nature of the many-body basis, ACE is inherently local or semilocal for graph ACE. Here, we employ descriptor-constrained density functional theory for parameterizing ACE with charge or other degrees of freedom, thereby transfering the variational property of the density functional to ACE. The descriptors can be of scalar, vectorial or tensorial nature. From the simplest case of scalar atomic descriptors we directly obtain charge-dependent ACE with long-range electrostatic interactions between variable charges. We observe that the variational properties of the charges greatly help in training, avoiding the need for charge-constrained DFT calculations.
\end{abstract}

\maketitle

% \section{Thoughts that we may want to move somewhere}

% Sometimes models that build on atomic charges $q_i$ only are considered point-charge models,  which is not correct as the atomic charges are obtained from a charge distribution but only the first moment of the charge distribution is then employed in the model.

% This scheme effectively overcomes the limitations inherent in the 4G-HDNNP approach~\cite{ko2021general,ko2021fourth,ko2023accurate}. In particular, the total energy is now ensured to be smooth and variational with respect to the atomic multipoles. 

% \rd{Need to complete references}

% \section{References that still need to be distributed}

% ACE from Goff with explicit derivation
% ~\cite{niklasson2021extended,niklasson2021extended_orbitalfree},

% constrained calculations

% keep:
% \cite{shapeev2016moment} 
% \cite{Drautz_PhysRevB.99.014104}
% \cite{batzner20223} \cite{musaelian2023learning} \cite{batatia2022mace} \cite{PhysRevResearch.4.L042019} 

\section{Introduction}

Charge transfer contributes to stabilize materials with ionic or polar bonds and varying charge states are critical for chemical reactions and transformations.  Atomistic theory typically starts from the Born-Oppenheimer approximation that implies that the energy or any other property of a many-atom system is parameterized as a function of the positions of the atoms. This means that the energy $E_{BO}(\br_1, \br_2, \dots, \br_N)$ or a suitably defined charge on atom $i$, $q_i(\br_1, \br_2, \dots, \br_N)$ depends solely on the positions $\br_1, \br_2, \dots, \br_N$ and species of the $N$ atoms in the system.  It therefore appears sensible to develop machine learning potentials that predict atomic charge as a function of atomic positions. This enables to estimate contributions of long range Coulomb interactions $q_i q_j/r_{ij}$ from models that use only local information of atoms surrounding atoms $i$ and $j$ to predict charges $q_i$ and $q_j$,   respectively. 

There are limitations to local models for atomic charges. First,  charge transfer in density functional theory is inherently non-local. The self-consistent variational state relates all atoms by a joint Fermi energy.  A small change to the system, for example, by the addition of an atom to a surface, can change the occupation of another state, for example,  at a defect atom, and with this the charge on the defect atom at an arbitrary distance from the atom on the surface. Only by including electronic temperature to broaden the population of localized states the relevance of non-local charge transfer diminishes. Second,  local models cannot comprise phase transitions that are unrelated or not directly related to atomic positions, such as, for example, charge ordering or ferromagnetic to paramagnetic transitions. Third, atomic charges or magnetic moments change with strength and direction of an external electric or magnetic field, which cannot be facilitated in a local model.

In the following we therefore assume that the energy is parameterized by atomic positions and charges simultaneously,
$E = E(\br_1, q_1, \br_2, q_2,\dots, \br_N,q_N) \,$.
In DFT this is realized by constrained optimization. While the parameters $\br_i$ and $q_i$ are kept fixed,  all other parameters are optimized to minimize the DFT functional\cite{PhysRevLett.53.2512, constrainedDFT,kaduk2012constrained}. As we will develop in detail in the following, the atomic charges $q_i$ are descriptors that allow one to model the charge state of an atom.  The energy in Born-Oppenheimer approximation is obtained by minimizing the expression for the energy with respect to the descriptors $q_i$,
\begin{equation}
E_{BO} = \text{Min!}_{{q_i}} E(\br_1, q_1, \br_2, q_2,\dots, \br_N,q_N) \,. \label{eq:BOopt}
\end{equation}
This approach is fully general and in different variations has been used for decades.  For example,  if one abstracts $q_i$ to vectors or tensors,  and takes them as electronic structure information,  such as the expansion coefficients of the basis functions in DFT, then the optimization in Eq.(\ref{eq:BOopt}) is nothing else but the variational optimization of the DFT functional.  This carries over to simplified models of the electronic structure, such as tight-binding models that are based on a second-order expansion of the DFT functional in a local orbital basis with orbitals $\alpha$ on atom $i$\cite{PhysRevB.39.12520,PhysRevB.58.7260, pettifor1995bonding,finnis2003interatomic,drautz2015bond,goff2023shadow,Thomas2024-sc}.  In tight binding models the orbital charges $q_{i \alpha}$ and onsite levels $E_{i \alpha} = \pder{E}{q_{i \alpha}}$ are remnants of the charge density $\rho(\br)$ and the effective potential $v_{eff} = \vder{E}{\rho}$ in DFT,  respectively. The second-order expansion of the DFT functional directly leads to expressions that are quadratic in charge.  If one further simplifies and groups orbital charges into atomic charges $q_i = \sum q_{i \alpha}$,  assumes  that the linear expansion coefficients and the diagonal quadratic expansion coefficients on the same atom are constant and that the off-diagonal second-order expansion coefficients between different  atoms $i$ and $j$ decay as $1/r_{ij}$, one obtains the Qeq model ~\cite{rappe1991charge,  mortier1986electronegativity} that is widely employed for introducing charge transfer into empirical potentials.

Therefore,  many models that incorporate variable charges can be understood and derived by approximations and simplifications of the DFT functional in two limiting cases.  First, charges on the Born-Oppenheimer energy surface that are parameterized as a function of atomic positions and second,  charges as independent variables that are formally obtained from constrained DFT calculations. Optimization of the energy with respect to these charges recovers the Born-Oppenheimer energy.  However,  this categorization is not always clearly followed in recent machine learning potentials literature.  For example,  some procedures mix self-consistently optimized charges with charges predicted from the local atomic environment, consider further descriptors for predicting the energy and charges, or invoke message passing to predict charges from a semi-local environment and a variety of different models and approaches are available today
~\cite{ghasemi2015interatomic,
faraji2017high,
grisafi2019incorporating,
metcalf2020electron,
xie2020incorporating,
unke2021spookynet,
ko2021general,
ko2021fourth,
khajehpasha2022cent2,
zhang2022deep,
cools2022modeling,
jacobson2022transferable,
staacke2022kernel,
ko2023accurate,
li2023long,
anstine2023machine,
huguenin2023physics,
deng2023chgnet}.

Here we devise a systematically extendable descriptor-based expansion suitable for incorporating charge and other degrees of freedom in machine learning interatomic potentials. We exploit the variational property of the density functional to ensure self-consistent model parameterization and demonstrate numerical implementation of charge transfer into the atomic cluster expansion (ACE).
 ACE~\cite{Drautz_PhysRevB.102.024104,Drautz_PhysRevB.99.014104} has proven to be a powerful and general tool to represent potential energy surfaces for different elements and alloys ~\cite{lysogorskiy2021performant,qamar2022atomic,ibrahim2023atomic,liang2023atomic,ibrahimEfficientParametrizationTransferable2024} featuring a complete basis of the atomic environment \cite{Drautz_PhysRevB.102.024104,Dusson22}. Furthermore, ACE is able to encompass additional degress of freedom, such as magnetic moments or charges~\cite{Drautz_PhysRevB.102.024104,rinaldi2024non} and semi-local interactions ~\cite{bochkarev2022multilayer,batatia2023general,batatia2022mace,Bochkarev2024GRACE}, tensor-reduced representation~\cite{darby2023tensor}, Hamiltonian matrices~\cite{zhang2022equivariant} and wave functions~\cite{drautz2022atomic}. Moreover, active learning based on uncertainty quantification can be naturally incorporated~\cite{van2023hyperactive,lysogorskiy2023active}. 
 %However, the possibility to fit charge-dependent models within PACEmaker~\cite{bochkarev_PhysRevMaterials.6.013804} is still missing. Furthermore, all the present flavors of ACE cannot distinguish between different charge states and do not include long-range interactions and charge transfer. In this context, we provide an implementation of the ACE+Qeq framework in PACEmaker and we benchmark it against other approaches. 
 
In Sec.~\ref{sec:basics} we provide the basic definition of atomic charges and dipoles in terms of the moments of the charge density and an alternative representation in terms of coefficients obtained by projecting the charge density on a basis. We then expand the energy in Sec.~\ref{sec:Charge and energy}. In the following Sec.~\ref{sec:Charge density constraints} and~\ref{sec:Charge constraints}, we approximate the variation of the charge density, as described by the expansion defined in the previous section, with a limited number of descriptors that act as constraints in the variational optimization and that are used to provide an expansion of the total energy in terms of few parameters, such as atomic charges and dipole moments. Details of the parameterization of the quadratic model based on charge constraints are provided in Sec.~\ref{sec:parameterization}. A study of the accuracy and stability achieved with the charge constraint model is provided in Sec.~\ref{sec:results}. In the final Sec.~\ref{sec:Discussions and conclusions} we conclude.

\section{Basics} \label{sec:basics}

Atomic or orbital charges are not uniquely defined. They vary depending on how they are computed and therefore the detailed prediction of atomic charges should not be a focus, but the accurate prediction of energies and forces as well as the electrostatic potential or electric field. We start from the charge density $\rho(\br)$ and assume a decomposition into atomic contributions
\begin{equation}
\rho(\br) = \sum_i^N  \rho_i(\br - \br_i) \,, \label{eq:rho}
\end{equation}
that are centered on the position of atom $i$, but do not specify the details of the decomposition.

It is customary to characterize charge densities by their moments $\int d\br \,\br \otimes   \br \otimes \dots  \otimes \br  \rho(\br)$.  The zeroth moment is the total charge of the $N$-atom system,
\begin{equation}
q = \sum_i^N q_i \, ,
\end{equation}
with the atomic charges
\begin{equation}
q_i = \int d\br  \rho_i(\br) \label{eq:atcharge}
\end{equation}
that follow from the decomposition of the charge density in Eq.(\ref{eq:rho}) and equivalently for the magnetization density
\begin{equation}
\pmb{M}_{i} = \int d\br \pmb{m}_{i} (\br),
\end{equation}
where $\pmb{m}(\br) = \sum_i^N  \pmb{m}_i(\br - \br_i)$.

The first moment of the charge density is given by
\begin{equation}
\pmb{\mu} = \sum_i  \br_i q_i + \pmb{\mu}_i 
\end{equation}
with the atomic dipole moments
\begin{equation}
\pmb{\mu}_i =  \int d\br \,   \br  \rho_i(\br) \,.
\end{equation} 
The second moment is the quadrupole tensor
\begin{equation}
\pmb{Q} =  \sum_i  \br_i  \otimes \br_i q_i + 2  \sum_i  \br_i  \otimes  \pmb{\mu}_i  +  \sum_i \pmb{Q}_i \,,
\end{equation}
with the atomic quadrupole moments
\begin{equation}
\pmb{Q}_i =  \int d\br \,   \br  \otimes \br \rho_i(\br) \,,
\end{equation} 
and so on for higher moments. Often the quadrupole tensor is given in traceless form $\pmb{Q}_i =  \int d\br \,   \left(\br  \otimes \br - \br^2 \pmb{1} \right) \rho_i(\br)$.  Because the tensor is symmetric,  in its traceless form it has only five independent entries.  Then the moments can also be obtained by integrating the density as $\int d\br r^l Y_{lm}(\hat{r})  \rho(\br)$,  with spherical harmonics $Y_{lm}$ and where $l=0$ corresponds to the zeroth moment,  $l=1$ is equivalent to the first moment,  $l=2$ to the second moment,  etc.

Instead of employing moments,  projection on orthogonal and complete local basis functions $\phib_{i\vi}(\br - \br_i)$
\begin{equation}
c_{i\vi} =  \int d\br  \phib_{i\vi}(\br - \br_i)  \rho(\br) \,,
\end{equation} 
enable a more flexible representation that also implies a separation of charge into atomic contributions
\begin{equation}
 \rho_i(\br ) = \sum_{\vi} c_{i\vi}   \phib_{i\vi}(\br) \,. \label{eq:rho1}
\end{equation} 
In particular for atomic properties local basis functions are useful,  $\phib_{i\vi}(\br - \br_i) = R_{nl}(r) Y_{lm}(\hat{r})$ with $\vi =nlm$,  $r = \| \br - \br_i \|$ and $\hat{r} =(\br - \br_i)/r$, and the local charge density
\begin{equation}
 \rho_i(\br ) = \sum_{\vi} c_{inlm} R_{nl}(r) Y_{lm}(\hat{r}) \,.\label{eq:rho2}
\end{equation} 
Summation over radial functions then recovers the moments,  for example,  for the zeroth moment,
\begin{equation}
q_i = \sum_n c_{in00} \,,
\end{equation}
or the dipole moment in spherical coordinates
\begin{equation}
d_{im} = \sum_n c_{in1m} \,,
\end{equation}
where we assumed suitable normalization of radial functions.

%\mr{if I plug Eq. 10 in Eq. 2, I obtain :
%\begin{equation*}
%q_{i}=\sum_{n}c_{in00}\int dr R_{n0}\left(|\br - \br_{i}|\right)r^{2}
%\end{equation*}
%is the integral equal to 1 for each n? I don't see radial functions normalization in the 2019 ACE paper
%}
%\mr{add how dipole moments can be defined in this notation: in spherical coordinates $\pmb{d}_{m}=\sum_{n}c_{in1m}$}

\section{Charge and energy} \label{sec:Charge and energy}

We start from a general local representation of the charge density as in Eqs.(\ref{eq:rho},\ref{eq:rho2}) and write the energy as 
\begin{equation}
E = E[\bR, \rho] =  E(\bR,\pmb{c}) \,,
\end{equation}
with $\bR = (\br_1, \br_2, \br_3, \dots)$ all atomic positions including chemical species information and $\pmb{c} = \{ c_{inlm} \}$, i.e. the charge density is fully characterized by the expansion coefficients $c_{inlm}$. The equilibrium charge density is obtained from optimization of $E - \mu (\sum_i q_i - Q)$, that couples conservation of the charge $Q$ in the system using the Lagrange multiplier $\mu$, i.e.
\begin{equation}
\pder{E}{c_{inlm}} = \mu \delta_{l0} \delta_{m0} \,. 
\end{equation}
This equation looks more familiar if written for atomic charges Eq.(\ref{eq:atcharge}) only, then
\begin{equation}
\pder{E}{q_{i}} = \mu \,,  
\end{equation}
with $\mu$ the electron chemical potential that is constant everywhere in equilibrium.

%The electrostatic energy is written as
%\begin{align}
%E^{\text{ES}} &= \frac{e^2}{2} \int  \frac{\rho(\br)\rho(\br')}{|\br -\br'|}\, d\br d\br' \nonumber \\
%&= \frac{1}{2} \sum_{\substack{inlm\\i'n'l'm'}} c_{inlm} c_{i'n'l'm'} f_{\substack{inlm\\i'n'l'm'}}(|\br_i -\br_{i'}|)\nonumber \\
%&= \frac{1}{2} \sum_{ii'} \pmb{c}_i \, \pmb{c}_{i'} \, \pmb{f}_{ii'}
%\end{align}
%where the last line is a shorthand notation and with
%\begin{equation}
%f_{\substack{inlm\\i'n'l'm'}}(\br_i -\br_{i'}) = e^2 \int  \frac{R_{nl}(r) Y_{lm}(\hat{r}) R_{n'l'}(r') Y_{l'm'}(\hat{r}') }{\br -\br'}\, d\br d\br' \,.\label{eq:f_coulomb}
%\end{equation}
%This expression becomes more intuitive when limited to spherically symmetric charges that are characterized by expansion coefficients with $l=m=0$, i.e., $c_{in00}$. Then
%\begin{equation}
%E^{\text{ES}} = \frac{1}{2} \sum_{ij} {q_i q_j} f_{ij}(| \br_i - \br_j|)  \,,\label{eq:es}
%\end{equation}
%\mr{this equation is correct only for n=1, where $q_{i}=c_{i100}$ and $q_{j}=c_{j100}$, change to $f_{ij}(\br_i - \br_j)$}
%where for large distances between atoms $i$ and $j$ a local decomposition of charge implies
%\mr{change to: for $| \br_i - \br_j|\rightarrow\infty$}
%\begin{equation}
% f(| \br_i - \br_j|) \propto \frac{1}{| \br_i - \br_j|} \,.
%\end{equation}
%\mr{change to $f(\br_i - \br_j) \rightarrow \frac{1}{| \br_i - \br_j|} \,$}

We expand the energy in terms of the expansion coefficients $\pmb{c}$,
\begin{align}
E &= E_0(\bR, \pmb{c}_0) + \pmb{A}(\bR, \pmb{c}_0) \Delta \pmb{c}+ \pmb{B}(\bR, \pmb{c}_0) \Delta \pmb{c} \Delta \pmb{c} \nonumber \\
&+ \pmb{C}(\bR, \pmb{c}_0) \Delta \pmb{c} \Delta \pmb{c}\Delta \pmb{c} + \dots \,, \label{eq:Eexp}
\end{align}
with $\Delta \pmb{c} = \pmb{c} - \pmb{c}_0$ and a suitably chosen reference $\pmb{c}_0$, for example, the charge density of overlapping charge neutral atoms. If one starts from a local decomposition of the one-electron wave function  $\psi_k$ of eigenstate $k$ in density functional theory, one obtains slightly different expressions and an expansion in the density matrix, not the charge density \cite{drautz2015bond,finnis2003interatomic,drautz2011}.
%Then the charge density is given by
%\begin{equation}
%\rho(\br) = \sum_k f_k | \psi_k(\br) |^2 \,,
%\end{equation}
%with occupation number $f_k$ and the expansion of the eigenstates
%\begin{equation}
%\psi_k = \sum_{inlm} a_{inlm} R_{nl}(r) Y_{lm}(\hat{r}) \,,
%\end{equation}
%results in an expansion of the energy in the density matrix and not the charge density \cite{XYZ}.

\section{Charge density constraints} \label{sec:Charge density constraints}

For the derivation of effective models of the atomic interaction we need avoid dealing with the variation of the complete charge density $\rho(\bR)$ (and magnetization density) as described by the expansion coefficients $\pmb{c}$, but we would like to describe the change of the charge density with only few descriptors. This is achieved by constraining the charge density (and magentization density) to target descriptor values, while minimizing the energy functional,
\begin{equation}
\vder{}{\rho} \left( E[\rho] + \sum_{iv} \lambda_{iv} \left[ \int \, d\br f_{iv}(\rho_i) - d_{iv} \right] \right) =0\,, \label{eq:rhocon}
\end{equation}
with Lagrange multipliers $\lambda_{iv}$, functions $f_{iv}(\rho_i)$ that evaluate the descriptors and $d_{iv}$ the numerical values of the descriptors. For example, for constraining charge on an atom from Eq.(\ref{eq:atcharge}) one has $f_{iq}(\rho_i)= \rho_i$ and $d_{iq} = q_i$. If in addition, one further constrains the dipole moments, then $f_{i\pmb{\mu}}(\rho_i)= \br \rho_i$ and $d_{i\pmb{\mu}} = \pmb{\mu}_i$ is coupled with additional Lagrange multipliers $\lambda_{i\pmb{\mu}}$, and so on for further constraints. The Lagrange multipliers can be viewed as constraining fields. If magnetization or polarization are constrained, the Lagrange multipliers are constraining magnetic or electric fields, respectively, and therefore our analysis immediately incorporates external fields.

Alternatively, as the expansion coefficients $\pmb{c}$ completely characterize the charge distribution, the constraints can also be expressed as a function of the expansion coefficients $\pmb{F}(\pmb{c}) = \int \, d\br \pmb{f}(\rho)$ and Eq.(\ref{eq:rhocon}) is written as
\begin{equation}
\pder{}{ \pmb{c}} \left( E(\bR, \pmb{c}) + \pmb{\lambda} \left[ \pmb{F}(\pmb{c}) - \pmb{d} \right] \right) =0\,.
\end{equation}
The manifold $\pmb{c}_0$ that fulfills this condition is parameterized by the values of descriptors $\pmb{d}$, $\pmb{c}_0 = \pmb{c}_0(\pmb{d})$ and so is the energy on the manifold,
\begin{equation}
E = E(\bR, \pmb{c}_0) = E(\bR, \pmb{d}) \,. 
\end{equation}
or charge density $\rho(\bR) = \rho(\bR, \pmb{d})$. 
%This allows us to rewrite Eq.(\ref{eq:Eexp}) in terms of descriptors
%\begin{align}
%E(\bR, \pmb{c}_0, \pmb{d})  &= E_0(\bR, \pmb{c}_0) + \pmb{A}_0(\bR, \pmb{c}_0) \pmb{d}+ \pmb{B}_0(\bR, \pmb{c}_0) \pmb{d} \pmb{d} \nonumber \\
%&+ \pmb{C}_0(\bR, \pmb{c}_0) \pmb{d} \pmb{d} \pmb{d} + \dots \,, \label{eq:EexpD}
%\end{align}
%\mr{does it make sense to say that $A_{0}(\bR, \pmb{c}_0)$ is equal to the product of $A(\bR, \pmb{c}_0)$ with $<\pmb{d}|\pmb{c}>$?}
%where $\pmb{c}_0$  that fulfills this condition is param $\pmb{c}_0 = \pmb{c}_0(\pmb{d})$

%which is analogous to Eq.(\ref{eq:rhocon}).
%\mr{is the partial derivative equal to $\pder{}{ \pmb{d}}$ instead of $\pder{}{ \pmb{c}}$?}
%This makes the charge density and energy are parameterized by the descriptor $\pmb{c}_0 = \pmb{c}_0(\pmb{d})$ or more explicitly $\rho(\bR) = \rho(\bR, \pmb{d})$ and $E = E(\bR, \pmb{d})$. 

Then applying expansion Eq.(\ref{eq:Eexp}) on the manifold results in
\begin{align}
E(\bR, \pmb{d})  &= E_0(\bR, \pmb{d}_0) + \pmb{\chi}(\bR,\pmb{d}_0) \Delta\pmb{d}+ \pmb{\eta}(\bR,\pmb{d}_0) \Delta\pmb{d} \Delta\pmb{d} \nonumber \\
&+ \pmb{\zeta} (\bR, \pmb{d}_0) \Delta\pmb{d} \Delta\pmb{d} \Delta\pmb{d} + \dots \,, \label{eq:EexpDD}
\end{align}
with $\Delta \pmb{d} = \pmb{d} -\pmb{d}_0$ and suitably chosen reference values of descriptors $\pmb{d}_0$. This formal expansion is suitable for the development of effective models as the energy is parameterized only by atomic positions, chemical elements and descriptor values. By restricting the expansion to constraint, optimized values of the charge density, the expansion ensures smooth variation of observables such as energy, forces, electrostatic potential or electric field. We note that the framework by Xie et al. \cite{xie2020incorporating} starts from an analogous analysis with the descriptors $\pmb{d}$ limited to charge populations $\pmb{p}$. We further comment that a series expansion in the form Eq.(\ref{eq:EexpDD}) may not be effective and more complex tensorial expansions may converge more efficiently \cite{Drautz_PhysRevB.102.024104}.

\section{Charge constraints} \label{sec:Charge constraints}

We limit the constraints to charges only, $d_{iq} = q_i$, which from Eq.(\ref{eq:EexpDD}) reads
\begin{align}
E(\bR, \pmb{q})  &= E_0(\bR) + \sum_i {\chi}_i(\bR) {q}_i+ \sum_{ij} {\eta}_{ij}(\bR) {q}_i {q}_j \nonumber \\
&+ \sum_{ijk} {\zeta}_{ijk} (\bR) {q}_i {q}_j {q}_k + \dots \,, \label{eq:EexpDq}
\end{align}
and where we took charge neutral atoms as reference and made the summation over atoms explicit. By construction the energy $E(\bR, \pmb{q})$ has a minimum when the charges take their unconstrained values $\pmb{q}_0$, i.e. then
\begin{equation}
\pder{}{ \pmb{q}} \left( E(\bR, \pmb{q}) + \mu(\sum q_i -Q) \right) = 0 \,,\label{eq:var_min}
\end{equation}
which we will exploit for the parameterization later. This also means that second order is the lowest possible order at which Eq.(\ref{eq:EexpDq}) can be terminated. If in addition to the charges,  dipole-moments are also constraint, in second order this generalizes to
\begin{align}
E &= E^{(0)}(\bR) + \sum_i \chi_i(\bR) q_i +  \sum_{i} \pmb{\chi}_{i\mu}(\bR) \pmb{\mu}_{i} \\
&+ \sum_{ij} \eta_{ij}(\bR) {q_i q_j} \nonumber + \sum_{ij} \pmb{K}_{ij}(\bR) q_i\pmb{\mu}_{j} \\ 
& + \sum_{ij} \pmb{\eta}_{ij\mu}(\bR) \pmb{\mu}_{i}\pmb{\mu}_{j}\,. \label{eq:sep4}
\end{align}

If the constraints are magnetic moments $\pmb{M}$ one obtains the Heisenberg-Landau energy expression often employed in spin-lattice dynamics, which to second order is given by
\begin{equation}
E = E^{(0)}(\bR) + \sum_i A_i(\bR) \pmb{M}_{i}^{2} + \sum_{ij}^{i \neq j} J_{ij}(\bR) \pmb{M}_{i} \pmb{M}_{j},
\end{equation}
and where without external field first order terms vanish due to time reversal symmetry.

\section{Parameterization} \label{sec:parameterization}

For the parameterization of the charge constraint model we assume locality for $ E_0(\bR)$,  $\chi_i(\bR)$ and $\eta_{ii}(\bR)$ in Eq.(\ref{eq:EexpDq}) and expand these quantities in terms of ACEs. We approximate the off-diagonal elements $\eta_{ij}(\bR)$ for $i \neq j$ to be of pure long-range electrostatic nature and to only depend on the distances $|\br_i - \br_i|$. The expression for overlapping atom-centered, Gaussian charge densities is given in Eq.(\ref{eq:a1}) of App.~\ref{appendix:matrixelements}. 

The DFT calculations that we used for training were self-consistent, i.e., we know that for a given structure $n$ the atomic charges $q_i^{(n)}$ fulfill $\pder{E^{(n)}}{q_i} = \mu^{(n)}$, which corresponds to $N^{(n)}-1$ equations for a structure with $N^{(n)}$ atoms. We request the same condition to be fulfilled by the charge constraint model, imposing that the Lagrange multiplier $\mu$ in Eq.(\ref{eq:var_min}) is a stationary point of the total constrained energy. To this end we employed the following loss function in training
\begin{align}
\mathcal{L} &= \sum_n w^{(E)}_n (E_n - E_n^{(\text{ref})})^2 \nonumber \\
            &+ \sum_n \sum_{i}^{N^{(n)}} w^{(F)}_{ni} (\pmb{F}_{ni} - \pmb{F}_{ni}^{(\text{ref})})^2 \nonumber \\
            % &+ \sum_n \sum_{i}^{N^{(n)}} w^{(q)}_{ni} ({q}_{ni} - {q}_{ni}^{(\text{ref})})^2 \nonumber \\ 
            &+ \sum_n \sum_{i}^{N^{(n)}} w^{(d_{v})}_{ni} ({d}_{niv} - {d}_{niv}^{(\text{ref})})^2 \nonumber \\ 
            &+ \sum_n \sum_{i}^{N^{(n)}} w^{(\text{eq})} ( \pder{E^{(n)}}{q_{ni}} - \mu^{(n)})^2 \nonumber \\ 
            &+ \sum_n  w^{(\text{tot})} ( \sum_{i}^{N^{(n)}} q_i^{(n)} - Q^{(n)})^2 \label{eq:loss}
\end{align}
with $\mu^{(n)} = \frac{1}{N^{(n)}} \sum_{i}^{N^{(n)}} \pder{E^{(n)}}{q_i}$. The first and second line match energies and forces, respectively. The third line matches descriptors, such as charges or dipoles, with reference $\textit{ab-initio}$ values. The weights will be denoted as $w^{(q)}_{ni}$ and $w^{(d)}_{ni}$ for charges and dipoles. As the definition of charges can be arbitrary to some extent, we only gave very small weights $w^{(q)}_{ni}$ to charge matching. In fact, reasonable charges can be obtained even with $w^{(q)}_{ni}=0$. The fourth line ensures that the charges are variational, i.e. they are in equilibrium as they were in equilibrium in the DFT reference calculations, too. The last line makes sure that the charge in each structure is conserved. As we limited our implementation to second order in charge, the last two lines in the loss function were imposed with the determination of atomic charges and Lagrange multipliers through the matrix inversion procedure illustrated in App.~\ref{appendix:matrixelements}.

\section{Results} \label{sec:results}

In this section we access the accuracy of the charge constraint model using the dataset by Ko et al.~\cite{ko2021fourth} and parts of the MD22 dataset by Chmiela et al~\cite{chmiela2023accurate}. The former covers a large variety of structures in different charge states, both non-periodic and periodic. The latter contains MD trajectories for biomolecules and supramolecules, where complex interactions, such as long-range electrostatics and dispersion, are involved. Finally, we will discuss the stability of the charge constraint model by running MD. 
% Finally, we will analyze the effect of truncated electrostatic contributions on error metrics and predicted charges. 

\subsection{Benchmark}

The dataset by Ko et al.~\cite{ko2021fourth} is divided into five sub-datasets comprising structures in different charge states (neutral, positive, and negatively charged), where charge-transfer mediated by long-range electrostatic interactions contributes significantly to the total energy. In particular, this dataset contains Ag trimers with total charge equal to $+1$ and $-1$, ionic Na$_{9}$Cl$_{8}$ clusters where a neutral Na atom is removed, carbon chains terminated with hydrogen atoms in the neutral and positively charged state and a periodic system composed of a gold cluster adsorbed on a MgO(001) slab. The last three cases involve situations where global charge-transfer and long-range electrostatic effects are preponderant and influence the equilibrium configuration of these structures. In Table~\ref{table:1} we report our error metrics in comparison with the values obtained with the 4G-HDNNP approach. In particular, we examine predictions obtained using a charge-independent model (indicated by the ACE column) and a charge constraint model. In the latter, we represent only $\chi_i$ and, separately, both $\chi_i$ and $\eta_{ii}$ through a local many-body expansion ($\chi$(ACE) and $\chi$(ACE)+$\eta$(ACE) columns, respectively). As done with the 4G-HDNNP framework, we also fit charges explicitly using the extended loss function of Eq.(\ref{eq:loss}) employing $\omega_{ni}^{(q)}$=1 and tuning $\omega_{n}^{(E)}$ and $\omega_{ni}^{(F)}$ to achieve better energy and forces metrics. Moreover, the short range energy $E^{(0)}\left(\pmb{R}\right)$ is represented in terms of two expansions embedded with the Finnis-Sinclair square root function~\cite{bochkarev_PhysRevMaterials.6.013804} and $\chi_{i}\left(\textbf{R}\right)$ and $\eta_{ii}\left(\textbf{R}\right)$ in terms of linear ACE expansions. These last two quantities can assume both positive and negative values. Therefore, for better stability during the learning procedure, we re-defined them as corrections to their corresponding free atom values $\chi_{i}^{0}$ and $\eta_{ii}^{0}$ as:
\begin{equation}
\chi_{i}\left(\pmb{R}\right) = \chi_{i}^{0} + \Delta\chi_{i}\left(\pmb{R}\right) \label{eq:chi0}
\end{equation}
and
\begin{equation}
\eta_{ii}\left(\pmb{R}\right) = \eta_{ii}^{0} + \Delta\eta_{ii}\left(\pmb{R}\right). \label{eq:J0}
\end{equation}
Finally, a cutoff of 6 $\angstrom$ was employed for each subset. \\
The large metrics for both energies and forces obtained with the charge-independent ACE in the case of $Ag_{3}^{+/-}$ are attributable to the degeneracies present in the sub-dataset. Specifically, there are configurations with the same atomic arrangements but in different total charge states that break the injectivity of the structure-representation map at any body order for a charge-independent model. The charge contstraint ACE+Q model lifts these degeneracies, as clearly shown in the second and third columns of Table~\ref{table:1}. Moreover, the fact that ACE+Q overall exhibits better metrics than the local ACE model suggests that the former is capable of describing systems significantly influenced by global interactions, such as electrostatics and non-negligible charge-transfer. The improvement in the metrics can be up to 70$\%$, as in the case of the force error obtained for the Na$_{8/9}$Cl$_{8}^{+}$ sub-dataset with the $\chi$(ACE)+$\eta$(ACE) model. Additionally, it is worth noticing that expressing $\eta$ with an ACE expansion yields a drastic improvement in the charge RMSE (up to two orders of magnitude in the case of Ag trimers), outperforming the metrics obtained with the 4G-HDNNP that uses charges as additional input to the neural networks used to describe the short-range energy. 

% spooky net ???????

\begin{table*}[t]
	\setlength{\tabcolsep}{15pt} % Default value: 6pt
    \renewcommand{\arraystretch}{1.0}
\centering
\begin{tabular}{|c l l l l l|}
\hline
             & & ACE & $\chi$(ACE) & $\chi$(ACE) + & 4G-HDNNP\\
             & & & & + $\eta$(ACE) &\\
\hline
    $Ag^{+/-}_{3}$ & E & 809.62 & 0.59 & $\textbf{0.21}$ & 1.323 \\
     & F & 285.81 & 23.67 & $\textbf{23.10}$ & 31.69\\
     & Q & - & 20.86 & $\textbf{0.415}$ & 9.976 \\
\hline
    $Na_{8/9}Cl_{8}^{+}$ & E & 1.55 & 0.57 & 0.71 & $\textbf{0.481}$ \\
     & F & 41.72 & 18.30 & $\textbf{12.35}$ & 32.78 \\
     & Q & - & 24.95 & $\textbf{7.02}$ & 15.83 \\
\hline
    $C_{10}H_{2}/C_{10}H_{3}^{+}$ & E & 0.76 & 0.87 & $\textbf{0.75}$ & 1.194 \\
     & F & 37.22 & 40.77 & $\textbf{35.16}$ & 78.00 \\
     & Q & - & 7.55 & $\textbf{2.00}$ & 6.577 \\
\hline
    $Au_{2}-MgO$ & E & 2.56 & 2.52 & 1.63 & $\textbf{0.219}$ \\
     & F & 88.70 & 56.51 & $\textbf{50.27}$ & 66.00 \\
     & Q & - & 8.84 & $\textbf{3.34}$ & 5.698 \\
\hline
\end{tabular}
    \caption{Test error metrics are reported for energies in meV/atom, forces in meV/Å, and charges in me for various ACE+Q flavors, with a comparison to the 4G-HDNNP values~\cite{ko2021fourth}.}
    \label{table:1}
\end{table*}

In Table~\ref{table:2} the error metrics for three subsets of the MD22 dataset are reported in comparison with the predictions obtained with the global symmetric gradient domain ML (sGDML) force field~\cite{chmiela2023accurate}. This framework is a global model, naturally incorporating global interactions such as dispersion and electrostatics. While van der Waals interactions mostly dominate the Buckyball catcher, long-range electrostatic effects are present in the cases of the Stachyose and Tetrapeptide datasets. As clearly visible from the table, similar metrics than with sGDML were achieved with charge constraint ACE+Q. This indicates that our method is able to describe complex long-range interactions in large molecules, such as those from the MD22 dataset. The ACE models were fitted with only $\chi$ expressed in terms of a local many-body expansion.

\begin{table*}[t]
	\setlength{\tabcolsep}{15pt} % Default value: 6pt
    \renewcommand{\arraystretch}{1.0}
\centering
\begin{tabular}{|c l l l|}
\hline
             &  & ACE+Q & sGDML\\
\hline
    Ac-Ala3-NHMe & E & $\textbf{0.39}$ & 0.40 \\
     & F & $\textbf{28.49}$ & 34\\
\hline
%     Fatty acid (DHA) & E &  &  &  &  1.0 \\
%      & F & &  &  & 33 \\
% \hline 
    Stachyose & E & $\textbf{1.80}$ & 2.0 \\
     & F & 48.24 & $\textbf{29}$ \\
% \hline     
%     DNA base pairs (AT-AT) & E & & & & 0.52 \\
%      & F & & & & 30 \\
% \hline
%     DNA base pairs (AT-AT-CG-CG) & E & & & & 0.52 \\
%      & F & & & & 31 \\
\hline
    Buckyball catcher & E & $\textbf{0.27}$ & 0.34 \\
     & F & 36.42 & $\textbf{29}$ \\
% \hline
%     Double-walled nanotube & E & & & & 0.47 \\
%      & F & & & & 23 \\
\hline
\end{tabular}
    \caption{Test error metrics are reported for energies in meV/atom, forces in meV/Å, and charges in me for various ACE+Q flavors, with a comparison to the sGDML values~\cite{chmiela2023accurate}. }
    \label{table:2}
\end{table*}

\begin{figure}[h]
    \centering
    \includegraphics[width=8.5cm]{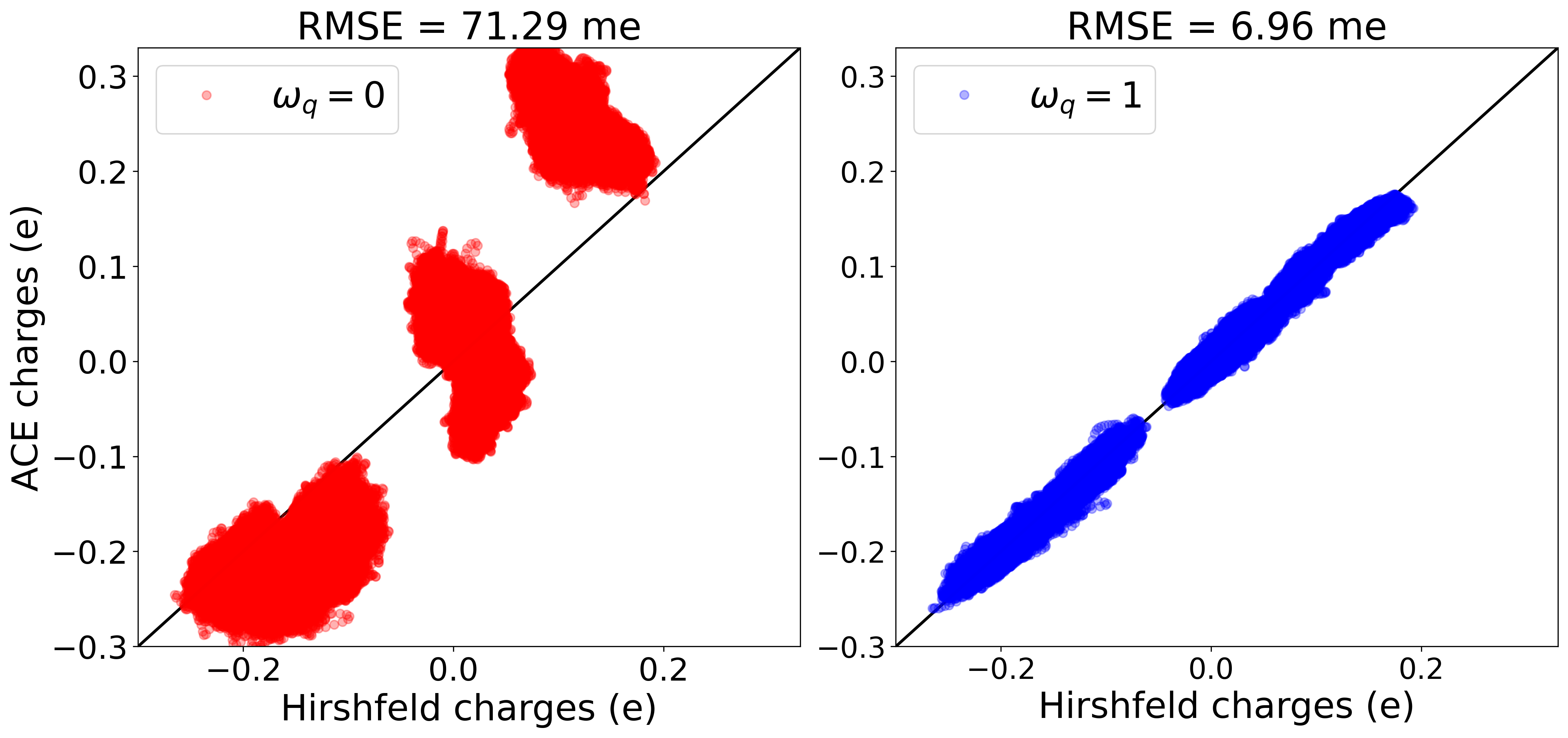}
    \caption{Correlation plots for predicted vs Hirshfeld charges for the test set of the Stachyose dataset. The fits were performed with $\omega_{ni}^{(q)}$=0 (left) and $\omega_{ni}^{(q)}$=1 (right). RMSE on charges are also reported.  RMSE on energy and forces are equal to 1.06 meV/atom and 65.32 meV/$\angstrom$ (left) and 1.35 meV/atom and 75.05 meV/$\angstrom$ (right).}
    \label{fig:corr_stc_chg}
\end{figure}

\begin{figure}[h]
    \centering
    \includegraphics[width=8.5cm]{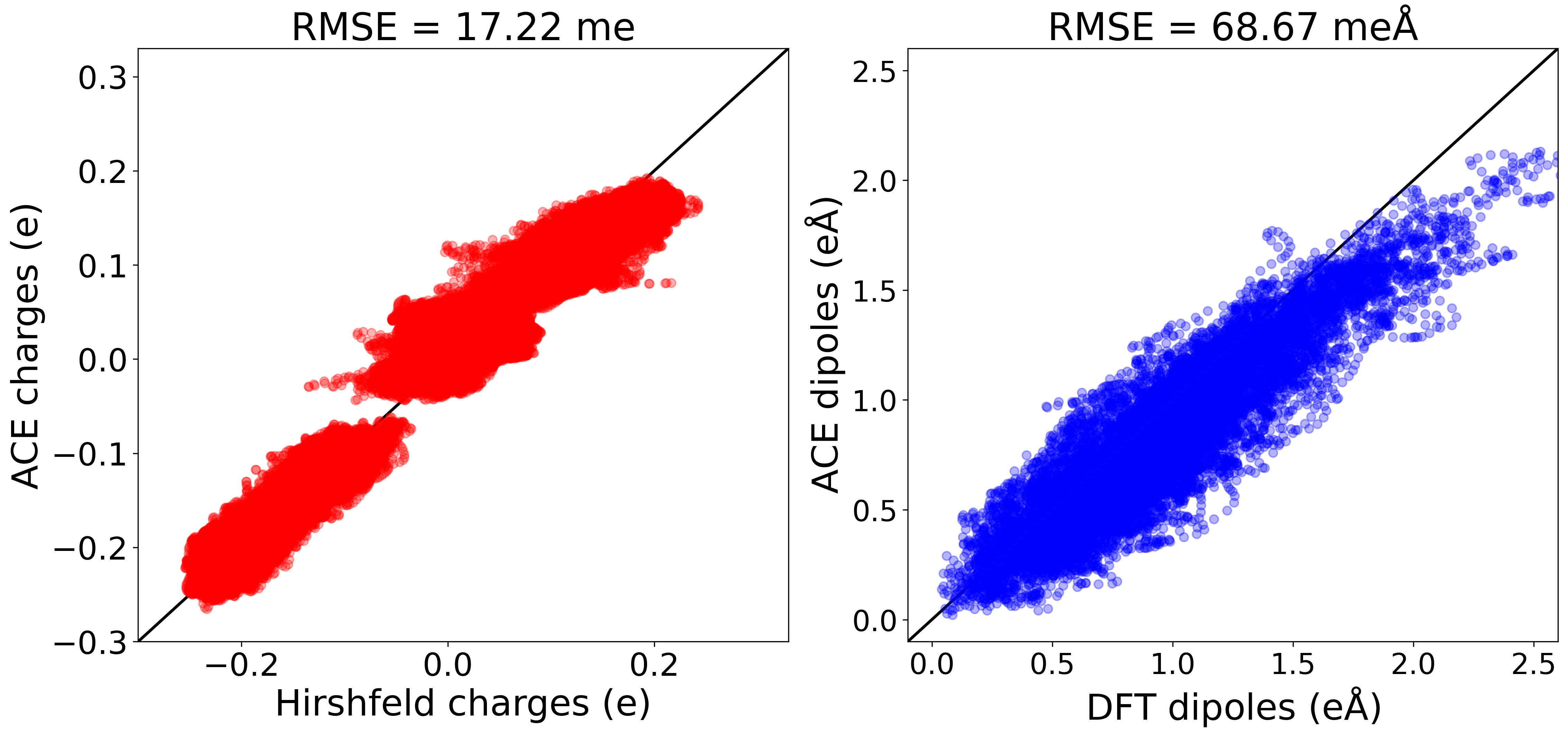}
    \caption{Correlation plots for charges (left) and total dipole moment (right) for the test set of the Stachyose dataset. The fits were performed with $\omega_{ni}^{(q)}$ and $\omega_{ni}^{(\mu)}$ equal to 1. RMSE on charges and total dipole moments are also reported. RMSE on energy and forces are equal to 1.61 meV/atom and 75.21  meV/$\angstrom$, respectively. }
    \label{fig:corr_stc_chg_dip}
\end{figure}

Finally, to check the predictions of the atomic charges and total dipole moments obtained with different values of $\omega_{ni}^{(q)}$ and $\omega_{ni}^{(\mu)}$ in Eq.~\ref{eq:loss}, we considered the Stachyose dataset and we compared the predictions with different combinations of the two weights, namely ($\omega_{ni}^{(q)}$=0, $\omega_{ni}^{(\mu)}$=0) and ($\omega_{ni}^{(q)}$=1, $\omega_{ni}^{(\mu)}$=0) in Fig.~\ref{fig:corr_stc_chg} and ($\omega_{ni}^{(q)}$=1, $\omega_{ni}^{(\mu)}$=1) in Fig.~\ref{fig:corr_stc_chg_dip}. The potentials were fitted using 1000 functions per element and only $\chi$ was expressed in terms of an ACE expansion. The calculation of the Hirshfeld charges and total dipole moments was performed with the FHIaims code~\cite{blum2009ab} using tight settings. As clearly visible from the left panel of Fig. ~\ref{fig:corr_stc_chg}, not fitting charges or dipoles at all results in reasonably predicted charges. Specifically, the oxygen atoms (i.e. negative values) exhibit the best correlation with the Hirshfeld reference charges. This agreement is primarily attributed to our choice of $\chi$ and $\eta$ in Eqs.(\ref{eq:chi0}) and (\ref{eq:J0}), where the corrections to the free atom values are expressed through ACE expansions. Explicitly fitting charges, as demonstrated in the right panel of the same figure, leads to a reduction in the charge error by one order of magnitude. In the case where both charges and total dipole moments are considered in the fit, the resulting error on the atomic charges is a bit larger than in the previous case, as shown in the left panel of Fig.~\ref{fig:corr_stc_chg}. This is attributed to the more complex loss function that needs to be minimized.

\subsection{Stability of charge constraint ACE+Q}

We demonstrate the stability of our approach by running molecular dynamics simulations in the NVT ensemble for the Stachyose system using the Berendsen thermostat from the ASE package~\cite{larsen2017atomic}. In this case, the potential was fitted representing also $\eta$ in terms of a local many-body expansion. Moreover, 1000 basis functions per element were considered during training with both weights $\omega_{ni}^{(q)}$ and $\omega_{ni}^{(\mu)}$ equal to zero. At every molecular dynamics (MD) iteration, the system of linear equations in Eq.(\ref{eq:matrixinv}) is solved before the force evaluation to obtain the instantaneous values of the atomic charges and Lagrange multiplier. Due to the cubic scaling involved in finding the equilibrium configuration of the atomic charges, shadow molecular dynamic schemes, such as the ACE+XL-QEq approach developed by Goff et al.~\cite{goff2023shadow}, which is based on extended Lagrangian (XL) Born-Oppenheimer molecular dynamics~\cite{niklasson2021extended}, could be employed to reduce the computational cost. However, here our focus is not to accelerate the relaxation of the atomic charges at each MD step, but rather to test the stability of our ACE+Q framework by monitoring the thermodynamic averages of quantities, such as $\chi_{i}$ and $\eta_{ii}$.

\begin{figure}[h]
    \centering
    \includegraphics[width=9.0cm]{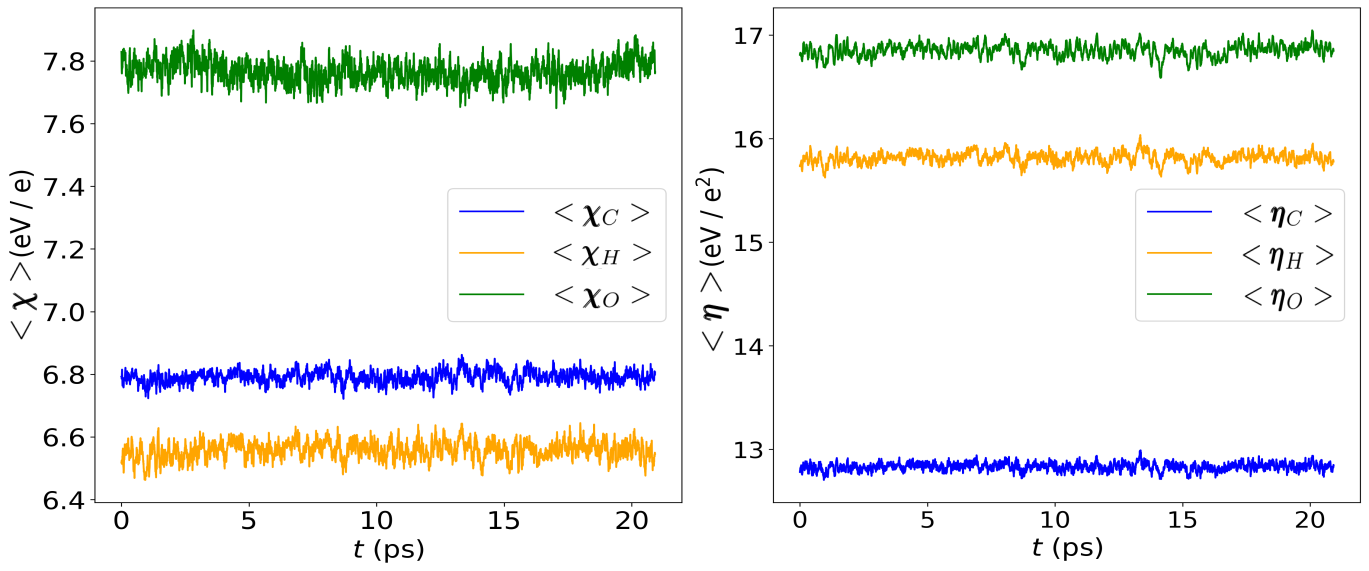}
    \caption{NVT dynamics for the Stachyose system. Left: variation with time of $\chi$ at the $C$, $H$ and $O$ sites. Right: variation with time of $\eta$ at the $C$, $H$ and $O$ sites.}
    \label{fig:stc_eq}
\end{figure}

In Fig.~\ref{fig:stc_eq} the running averages for atomic $<\chi>$ and $<\eta>$ are plotted for the different chemical species involved, namely C, O and H, for 20 ps of dynamics. As demonstrated by these plots, there are no unphysical jumps present during the time evolution. Therefore, the resulting charges predicted via charge equilibration  have stable trajectories. Furthermore, the fluctuations of $<\eta>$ over time are smaller in magnitude than those of $<\chi>$. This observation aligns with expectations, considering that $\chi$ represents the coefficient of the first order in the Taylor expansion. The evaluation time for a single MD step with the ASE calculator is equal to 2 ms$/$atom.  

% $\chi_{0}^{C}$=5.343, $\chi_{0}^{O}$=8.741, $\chi_{0}^{H}$=4.528
% $J_{0}^{C}$=10.126, $\chi_{0}^{O}$=13.364, $\chi_{0}^{H}$=13.8904

% pg_qeq_md_stc_j.py

\section{Discussion and conclusions} \label{sec:Discussions and conclusions}

% Sometimes models that build on atomic charges $q_i$ only are considered point-charge models,  which is not correct as the atomic charges are obtained from a charge distribution but only the first moment of the charge distribution is then employed in the model.

% \rd{Needs work}
We introduced a general framework, based on descriptor-constrained DFT, that generalizes previous ML potentials limited to a second order expansion in the atomic charges. The constraints that enter the modified variational problem can be of scalar, vectorial and tensorial nature. The energy is obtained as the variationally optimized DFT functional constrained to descriptor values, which in turn means that the energy is parameterized by the descriptors, including contributions from external fields. Series expansion of the energy in terms of the descriptors immediately recovers frequently used models, such as Qeq for charge transfer or the Heisenberg model of magnetic interactions.

In order to test our methodology and implementation, we parameterized a lowest order self-consistent charge constraint model with ACE. In this approach the short-range charge-independent energy, the first and second order onsite coefficients, are represented in terms of local many-body expansions. Equilibrium charges are obtained by minimizing the energy of the corresponding coarse-grained DFT constrained optimization problem. The resulting second-order/lowest-order self-consistent charge constraint ACE+Q model is closely related to second-order expansions of the DFT functional in the tight-binding approximation. Alternatively, it may be seen as a generalization of the Qeq framework to fully environmentally dependent first and second order onsite terms $\chi$ and $\eta$. In contrast to the 4G-HDNNP approach by Ko et al.~\cite{ko2021fourth,ko2021fourth,ko2023accurate} and similar to Goff et al.~\cite{goff2023shadow}, the ACE+Q energy contributes directly to the total energy and ensures that chargers are variational. Specifically, the charges used to calculate the energy correspond to charges of the instantaneous electronic ground state. We noted in our numerical test that the variational condition of the equilibrium charges were oftentimes sufficient for the parameterization, while fitting to explicit constraint charge values was not required. This is particularly useful as atomic charges are non-unique and defined in different ways. In addition this observation further suggests charge constrained DFT calculations may not be necessary for training.

Our ACE+Q model was implemented in the PACEmaker~\cite{bochkarev_PhysRevMaterials.6.013804} code and integrated into the ACE ecosystem. We further demonstrated that ACE+Q can achieve superior overall metrics compared to the 4G-HDNNP approach and similar metrics to global models like sGDML. Furthermore, the ACE+Q simplicity and physical transparency leads to stable molecular dynamics, with atomic charges evolving at each time step. This capability opens the door for large-scale MD simulations with ACE that incorporate charge-transfer and long-range electrostatic effects. In the future we plan to couple ACE+Q to shadow molecular dynamics \cite{goff2023shadow} to speed-up the calculation of the equilibrium charges at each MD step. Additionally, we will extend the ACE+Q expansion to higher order or use non-linear expansions in the charges.

\section*{Acknowledgments}

Funding by the Deutsche Forschungsgemeinschaft (DFG) in the SFB1394 Structural and chemical atomic complexity – from defect phase diagrams to material properties (Project ID 409476157) is gratefully acknowledged.

\section{Appendix}

\newpage

\appendix
\section{Connection to the Qeq model and solution of the constrained variational problem}
\label{appendix:matrixelements}
% \rd{Needs update}

Starting from the general expansion of the total energy of Eq.(\ref{eq:EexpDq}), one can directly simplify to the familiar Qeq expression used extensively in other work~\cite{ghasemi2015interatomic,faraji2017high,ko2021general,ko2021fourth,ko2023accurate,staacke2022kernel,jacobson2022transferable} that in the non-periodic case reads,
\begin{equation}
\eta_{ij}=\frac{1}{2}\left(J_{i}+\frac{1}{\sigma_{i}\sqrt{\pi}}\right)\delta_{ij} + \frac{1}{2}\frac{\erf\left(\frac{r_{ij}}{\sqrt{2}\gamma_{ij}}\right)}{r_{ij}}\left(1-\delta_{ij}\right).\label{eq:a1}
\end{equation}
In this simplified form, the diagonal contributions originate from the expansion of both short- and long-range energy, while the off-diagonal ones are of long-range nature. The quantity $J_{i}$, often termed as atomic hardness, is purely local. The remaining contributions are obtained from the electrostatic energy, where for the Coulomb integral the radial functions are taken as atom-centered Gaussians. Moreover, $\sigma_{i}$ corresponding to the width of these Gaussian functions and,
\begin{equation}
\gamma_{ij}=\sqrt{\sigma_{i}^{2}+\sigma_{j}^{2}}.
\end{equation}
In our implementation, $\sigma_{i}$ is set equal to the reference covalent radii of the atom $i$. \\
The solution of the charge constrained variational problem can be obtained by minimizing the total constrained energy with respect to the atomic charges and maximizing its value with respect to the Lagrange multiplier. These conditions are equivalent to the solution of the following system of linear equations:
\begin{equation}\label{eq:linsys}
\begin{cases}
\pder{}{ \pmb{q}} \left( E(\bR, \pmb{q}) + \mu(\sum q_i -Q) \right) = 0  \\
\sum_{i}q_{i} - Q = 0 
% \pder{}{ \mu }\left( E(\bR, \pmb{q}) + \mu(\sum q_i -Q) \right) = 0
\end{cases}.
\end{equation}
This $\left(N+1\right)\times\left(N+1\right)$ system of linear equations can be cast in the following matrix form:
\begin{equation}\label{eq:matrixinv}
\left(\begin{array}{c c c c c |c} 
     & & & & & 1\\ 
     & & & & & \\ 
     & & \textbf{A} & & & \vdots\\ 
     & & & & & \\ 
	 & & & & & 1\\ 
	\hline 
	1 &  & \hdots &  & 1 & 0\\  
\end{array}\right)
\begin{pmatrix}
       q_{1} \\
        \\
        \vdots\\
        \\
        q_{N} \\
       \hline 
       \mu \\
\end{pmatrix}
=\begin{pmatrix}
       -\chi_{1} \\
        \\
        \vdots\\
        \\
        -\chi_{N} \\
       \hline 
       Q_{tot}, \\
\end{pmatrix}
\end{equation}
where the matrix $\textbf{A}$ contains the electrostatic energy and atomic hardness contributions. A unique solution to the linear system of Eq.(\ref{eq:linsys}) can be achieved only if the matrix $\textbf{A}$ is positive definite. This is fulfilled only if $J_{i}>0$ for every atom $i$. The values of the atomic charges and Lagrange multiplier can then be obtained by matrix multiplication of the inverted  matrix $\textbf{A}$ and the column vector on the right-hand side of Eq.~\ref{eq:matrixinv}.

In non-periodic systems, the matrix $\textbf{A}$ in Eq.(\ref{eq:matrixinv}) is given by,
\begin{equation}
A_{ij} = \begin{cases}
      J_{i} + \frac{1}{\sigma_{i}\sqrt{\pi}} & \text{if $i=j$} \\
      \frac{\erf{\left(\frac{r_{ij}}{\sqrt{2}\gamma_{ij}}\right)}}{r_{ij}}, & \text{otherwise},
    \end{cases}.  
\end{equation}\\
The periodic case is handled with the Ewald summation~\cite{ewald1921berechnung,lee2009ewald}, where the scaling with the number of atoms is $N^{\frac{3}{2}}$. The resulting matrix elements have real-space contributions that read as:
\begin{equation}
A_{ij} = \begin{cases}
      J_{i} -\sqrt{\frac{2}{\pi}}\frac{1}{\eta} + \frac{1}{\sigma_{i}\sqrt{\pi}} & \text{if $i=j$} \\
      \frac{\erfc\left({\frac{r_{ij}}{\sqrt{2}\eta}}\right)}{r_{ij}} - \frac{\erfc\left({\frac{r_{ij}}{\sqrt{2}\sigma_{i}}}\right)}{r_{ij}}, & \text{otherwise},
    \end{cases}  
\end{equation}
where $\eta$ is the width of the Gaussian charges used in the Ewald procedure. This real-space contribution comprises alsocorrections to the point charge approximation used in the conventional Ewald sum~\cite{ewald1921berechnung,lee2009ewald,gingrich2010ewald}. The additional reciprocal space contribution reads,
\begin{equation}\label{eq:rec}
A_{ij} = \frac{4\pi}{V}\sum_{\textbf{k}\neq\textbf{0}}\frac{\exp\left(-\frac{\eta^{2}\textbf{k}^{2}}{2}\right)}{\textbf{k}^{2}}\cos\left[\textbf{k}\cdot\left(\textbf{r}_{i}-\textbf{r}_{j}\right)\right],
\end{equation}
where the indices $i$ and $j$ indicate all the possible combinations of pairs of atoms in the supercell. The computational cost of this contribution can be drastically reduced employing the symmetry of the cosine function. For instance, if $\textbf{k}$ is given in terms of triplets $k_{1}$, $k_{2}$ and $k_{3}$ of integers as: $\textbf{k}=k_{1}\textbf{b}_{1}+k_{2}\textbf{b}_{2}+k_{3}\textbf{b}_{3}$, where $\textbf{b}_{1}$, $\textbf{b}_{2}$ and $\textbf{b}_{3}$ are reciprocal lattice vectors, then, if $k_{1}$, $k_{2}$ and $k_{3}$ are not zero, the expression~\ref{eq:rec} can be simplified to:
\begin{equation}\label{eq:rec}
\begin{split}
&\sum_{\textbf{k}\neq\textbf{0}}\cos\left[\textbf{k}\cdot\left(\textbf{r}_{i}-\textbf{r}_{j}\right)\right] \\ &= 8\sum_{k_{1}k_{2}k_{3}}\cos\left(k_{1}r_{ij,x}\right)\cos\left(k_{2}r_{ij,y}\right)\cos\left(k_{3}r_{ij,z}\right),
\end{split}
\end{equation}
with $k_{1}>0$, $k_{2}>0$ and $k_{3}>0$.

% \section{Matrix elements in the case of truncated electrostatic interactions}
% \label{appendix:matrixelements_trunc}
% In the case of Wolf summation~\cite{wolf1999exact} the matrix elements are given by:
% \begin{equation}
% A_{ij} = \begin{cases}
%       J_{i} -\left[\frac{\erfc\left(\alpha r_{c}\right)}{r_{c}} + \frac{2\alpha}{\sqrt{\pi}}\right] + \frac{1}{\sigma_{i}\sqrt{\pi}} & \text{if $i=j$} \\
%       \left[\frac{\erfc\left(\alpha r_{ij}\right)}{r_{ij}} - 
%       \frac{\erfc\left(\alpha r_{c}\right)}{r_{c}}\right] -
%       \frac{\erfc\left({\frac{r_{ij}}{\sqrt{2}\sigma_{i}}}\right)}{r_{ij}}, & \text{otherwise},
%     \end{cases}  
% \end{equation}
% The more complex case of DSF~\cite{fennell2006ewald} method gives diagonal matrix elements as follows:
% \begin{equation}
% A_{ii} = J_{i} -\left[\frac{\erfc\left(\alpha r_{c}\right)}{r_{c}} + \frac{2\alpha}{\sqrt{\pi}} - \phi\left(r_{c}\right)r_{c}\right] + \frac{1}{\sigma_{i}\sqrt{\pi}}
% \end{equation}
% and off-diagonal elements:
% \begin{equation}
% \begin{split}
% A_{i\neq j} =  & \left[\frac{\erfc\left(\alpha r_{ij}\right)}{r_{ij}} - \frac{\erfc\left(\alpha r_{c}\right)}{r_{c}} - \phi\left(r_{c}\right)\left(r_{ij}-r_{c}\right)\right] - \\ & + \frac{\erfc\left({\frac{r_{ij}}{\sqrt{2}\sigma_{i}}}\right)}{r_{ij}}
% \end{split}
% \end{equation}
% where $\phi\left(r_{c}\right)$ is given by:
% \begin{equation}
% \phi\left(r_{c}\right) = -\frac{2\alpha}{\sqrt{\pi}}\frac{\exp(-\alpha^{2}r_{c}^{2})}{r_{c}}-\frac{\erfc\left(\alpha r_{c}\right)}{r_{c}^{2}}
% \end{equation}

%\bibliographystyle{unsrt}
%\bibliography{bibliography}

\end{document}